\documentclass[prl,aps, amssymb,amsmath,reprint,superscriptaddress,showpacs]{revtex4-1}
\pdfoutput=1

\usepackage[T1]{fontenc} 
\usepackage[utf8]{inputenc} 
\usepackage[UKenglish]{babel} 
\usepackage{hyperref}

\usepackage{color}
\usepackage{booktabs} 
\usepackage{array} 
\usepackage{graphicx}

\newcommand{\ket}[1]{|#1\rangle}

\newcommand{\supp}[0]{Note1}

\newcommand{\dsum}[0]{\displaystyle\sum}

\begin{document}

\author{M.E. Schwartz}
\affiliation{Quantum Nanoelectronics Laboratory, Department of Physics, University of California, Berkeley, California 94720, USA.}
\author{L. Martin}
\affiliation{Quantum Nanoelectronics Laboratory, Department of Physics, University of California, Berkeley, California 94720, USA.}
\author{E. Flurin}
\affiliation{Quantum Nanoelectronics Laboratory, Department of Physics, University of California, Berkeley, California 94720, USA.}
\author{C. Aron}
\affiliation{ Department of Electrical Engineering, Princeton University, Princeton, NJ 08544, USA.}
\affiliation{Laboratoire de Physique Th\'eorique, \'Ecole Normale Sup\'erieure, CNRS, Paris, France.}
\affiliation{Instituut voor Theoretische Fysica, KU Leuven, Belgium.}
\author{M. Kulkarni}
\affiliation{Department of Physics, New York City College of Technology, The City University of New York, Brooklyn, NY 11201, USA}
\author{H.E. Tureci}
\affiliation{ Department of Electrical Engineering, Princeton University, Princeton, NJ 08544, USA.}
\author{I. Siddiqi}
\affiliation{Quantum Nanoelectronics Laboratory, Department of Physics, University of California, Berkeley, California 94720, USA.}

\title{Stabilizing entanglement via symmetry-selective bath engineering in superconducting qubits}


\begin{abstract}
Bath engineering, which utilizes coupling to lossy modes in a quantum system to generate non-trivial steady states, is a tantalizing alternative to gate- and measurement-based quantum science.  Here, we demonstrate dissipative stabilization of entanglement between two superconducting transmon qubits in a symmetry-selective manner.  We utilize the engineered symmetries of the dissipative environment to stabilize a target Bell state; we further demonstrate suppression of the Bell state of opposite symmetry due to parity selection rules.  This implementation is resource-efficient, achieves a steady-state fidelity $\mathcal{F} = 0.70$, and is scalable to multiple qubits.

\end{abstract}

\date{\today}

\maketitle

Advances in quantum circuit engineering \cite{Koch2007, Manucharyan2009, Paik2011, Barends2013} have enabled coherent control of multiple long-lived qubits based on superconducting Josephson junctions \cite{Barends2014, Corcoles2015, Riste2015}.  Conventional approaches for further boosting coherence involve minimizing coupling to lossy environmental modes, but this poses an increasingly non-trivial challenge as chip designs scale and increase in complexity.  An alternate approach, quantum bath engineering \cite{Poyatos1996, Lin2013, Krauter2011, Barreiro2011}, explicitly utilizes this coupling in conjunction with microwave drives, to modify the dissipative environment and dynamically cool to a desired quantum state. Bath engineering in superconducting qubits has resulted in the stabilization of a single qubit on the Bloch sphere \cite{Murch2012}, a Bell-state of two qubits housed in the same cavity \cite{Shankar2013}, many-body states \cite{Hacohen-Gourgy2015}, and a variety of non-classical resonator states \cite{Leghtas2015, Holland2015}.  Additionally, theoretical proposals have been put forward for dissipative error correction \cite{Cohen2014, Kapit2015, Kerckhoff2010} and ultimately a universal quantum computation \cite{Verstraete2009, Mirrahimi2014}. 

These approaches require careful selection of the bath modes, and typically many drives to excite these modes so as to produce a non-trivial ground state.  Bath engineering schemes have typically focused on sculpting a density of states conducive to cooling, relying on the conservation of energy between drive, qubit, and resonator modes in multi-photon processes.  In this Letter, we harness an additional degree of freedom: the spatial symmetry of the bath, which mandates conservation of parity.  We combine both spectral and symmetry selectivity of the bath to provide a scalable protocol for generating on-demand entanglement using only a single microwave drive with a controllable spatial profile.  As a demonstration of this scheme, we generate and stabilize a two-qubit entangled state of choice in the single-excitation subspace using two tunable 3D transmon qubits \cite{Paik2011} in independent microwave cavities.  Our results demonstrate the viability of this protocol for stabilizing many-body entangled states with high fidelity in extended arrays.

\begin{figure}[ht]
\includegraphics[width=8.6cm]{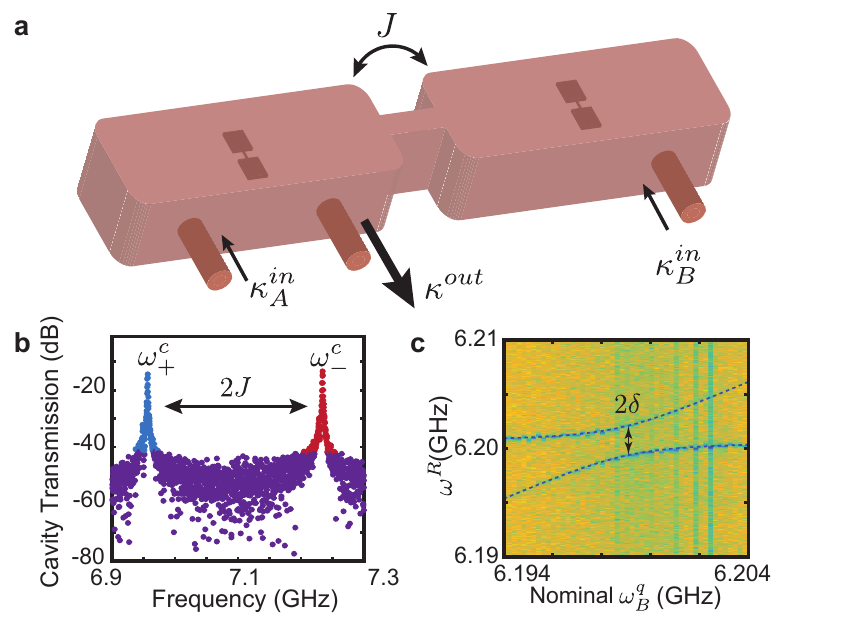}
\caption{Cavity-mediated qubit coupling. \textbf{a}:  schematic of aperture-coupled cavities, with weakly-coupled input ports $\kappa^{in}_i$, strongly-coupled output port $\kappa^{out}$, and inter-cavity coupling $J$. \textbf{b}: Transmission spectrum of the coupled cavity modes, showing the symmetric (blue) and antisymmetric (red) peaks. \textbf{c}: Pump-probe spectroscopy of the coupled qubit modes, exhibiting an avoided crossing.  Cavity $B$ is driven at the symmetric cavity resonance conditioned on the qubit state $|gg\rangle$, and cavity $A$ is driven at a swept Rabi pump frequency $\omega^R$.  A dip in transmission (blue) indicates that $\omega^R$ is resonant with a qubit mode.  Vertical stripes are an artifact of data acquisition.  The dashed line is a fit of the spectral data.}
\label{fig1}
\end{figure}

The experiments are implemented (Figure \ref{fig1}a) using two aperture-coupled copper waveguide cavities (indexed as $A$ and $B$ throughout this Letter), with an independent flux-tunable transmon embedded in each cavity.  The cavities are fabricated with near-identical resonance frequencies $\omega^c_{A,B}  \equiv \omega^c = 2\pi \times 7.114 \text{ GHz}$; the qubits are flux-tuned to resonance at $\omega^q_{A,B} \equiv \omega^q = 2\pi \times 6.200 \text{ GHz}$.  The full set of qubit and cavity parameters are tabulated in the Supplemental Material \footnote{See supplementary information.}.  The cavities are  individually addressable via a weakly-coupled port ($\kappa^{in}_i$) through which we apply qubit pulses and bath drives; cavity $A$ has an additional strongly coupled port for readout. 

The unitary dynamics of the system are described by a Hamiltonian that can be subdivided into qubit, cavity, and drive components: 
\begin{equation}
\hat{\mathcal{H}} = \hat{\mathcal{H}}_q + \hat{\mathcal{H}}_a + \hat{\mathcal{H}}_d
\end{equation}
where, in the rotating wave approximation,
\begin{equation}
\begin{array}{lcl}
\hat{\mathcal{H}}_q & = & \dsum\limits_{i=A,B}\left[\dfrac{\omega^q}{2}\hat{\sigma}^z_i + g_i\left(\hat{\sigma}^+_i \hat{a}_i + \hat{\sigma}^-_i \hat{a}^\dagger_i\right) \right]\\
\hat{\mathcal{H}}_a & = &  \dsum\limits_{i=A,B} \left[ \omega^c\hat{a}^\dagger_i \hat{a}_i + J\left(\hat{a}_A \hat{a}_B^\dagger + \hat{a}_A^\dagger \hat{a}_B\right) \right] \\
\hat{\mathcal{H}}_d & = & \dsum \limits_{i=A,B}  \epsilon^d_i  \left[ \hat{a}_i^\dagger e^{-i(\omega^d t +\phi_i)} +\hat{a}_i e^{i(\omega^d t +\phi_i)}\right] \\
\end{array}
\end{equation}
Here,  $\hat{\sigma}_i$ are Pauli operators on the qubits; $\hat{a}_i^\dagger$ are creation operators in the cavity modes; $\epsilon^d_i$ are Rabi drives applied at a single frequency $\omega^d$ to the respective cavities with a tunable phase $\phi_i$; and $g_i$ are the qubit-cavity couplings.  Decay mechanisms not accounted for in these unitary dynamics include qubit energy relaxation ($\Gamma_1$) and dephasing ($\Gamma_\phi$), and cavity photon leakage ($\kappa$).

The effects of the coupling terms $g$ and $J$ manifest in both the qubit and cavity sectors.  The central cavity resonances hybridize into symmetric and antisymmetric modes, with the former having a lower frequency (Figure \ref{fig1}b). We define these modes as $\omega^c_\pm = \omega^c \mp J$. In the dispersive limit where the qubit-cavity detuning $\Delta_\pm \equiv \omega^q -\omega^c_\pm$ is large in comparison to $g$, the qubit-cavity coupling creates a photon-mediated $XY$ interaction between the qubits, lifting the degeneracy in the single-excitation subspace \cite{Aron2014a}. Defining $\delta = J\dfrac{g_Ag_B}{\Delta_+\Delta_-}$, the coupled eigenstates and eigenenergies are given by the following:

\begin{equation}
\begin{array}{lclclcl}
\ket{T_-} &\equiv& \ket{gg}&;& \omega_{\ket{T_-}} &=& 0 \\
\ket{T_0} &\equiv &\dfrac{\ket{ge} + \ket{eg}}{\sqrt{2}} & ; & \omega_{\ket{T_0}}& =& \omega^q - \delta \\
\ket{S} &\equiv& \dfrac{\ket{ge} - \ket{eg}}{\sqrt{2}} & ; & \omega_{\ket{S}} &= &\omega^q + \delta \\
\ket{T_+} &\equiv& \ket{ee}&;& \omega_{\ket{T_+}} &=& 2\omega^q \\
\end{array}
\end{equation}
Note that the single-excitation states $\ket{S}$ and $\ket{T_0}$ are maximally-entangled Bell states.  We can then define full basis states of the system including the cavity modes, as 
\begin{equation}
 \ket{i,j,k} = \ket{n_+} \otimes \ket{n_-} \otimes \ket{\psi_q}
\end{equation}
where $n_\pm$ indexes the Fock state of the respective hybridized cavity modes and $\ket{\psi_q}$ is a coupled qubit state $\ket{\psi_q} \in \{\ket{S}, \ket{T_{0,\pm}} \}$.  Figure \ref{fig1}c shows the qubit-sector avoided crossing of width $2\delta = 2\pi \times 2.7$ MHz, in quantitative agreement with independently-characterized system parameters.  

\begin{figure}
\includegraphics[width=8.6cm]{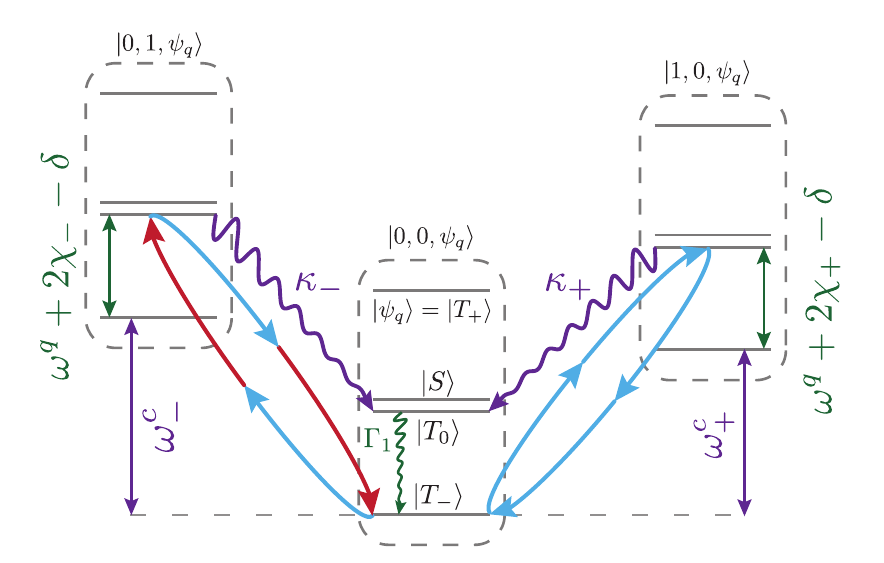}
\caption{Protocol for cooling to $\ket{0,0,T_0}$ via $\omega^c_-$ (left) and $\omega^c_+$ (right).  Each set of levels outlined in grey dashed lines represents a rung on the Jaynes-Cummings ladder; the states $\ket{\psi}$ are the coupled qubit states.  Not drawn to scale. The illustrated drives (arrows) represent $\omega^d_{\ket{T_0}}(\pm)$ from Equation \ref{eq:freqs}. Symmetry selection rules require that if cooling via $\omega^c_+$, the drive must be overall symmetric (indicated by blue lines), with $\phi = \{0, \pi\}$; if cooling via $\omega^c_-$, the drive must comprise one antisymmetric (red) photon for each symmetric photon ($\overline{n}_+ = \overline{n}_-$). If this condition is met, stochastic leakage of cavity photons (purple, $\kappa$) brings the system to the entangled state $\ket{0,0,T_0}$.  Leakage from the entangled state is dominated by qubit decay (green, $\Gamma_1$); leakage rates not shown include transitions between $\ket{S}$ and $\ket{T_0}$, and off-resonant pumping to $\ket{T_+}$.  }
\label{fig:cool}
\end{figure}

We now aim to stabilize the entangled state of choice ($\ket{S}$ or $\ket{T_0}$) by taking advantage of the distinct symmetries of the bath modes at $\omega^c_+$ and $\omega^c_-$ \footnote{Since the target entangled states are in fact eigenstates of the coupled Hamiltonian, it is in principle possible to coherently pulse to these states.  However, because the splitting is small, a coherent pulse with narrow enough bandwidth to drive selectively to one of these states would need to be several microseconds long, and therefore would be spoiled by qubit decay.}.
  We do this by simultaneously applying a two-photon at the individual cavity ports while varying the relative phase between the cavities (Figure \ref{fig:cool}).  This work represents a generalization to arbitrary drive phase of the proposal in \cite{Aron2014a}; the full theoretical treatment of this generalization (including dynamics) is presented in the Supplement \cite{\supp}.  

Our cooling protocol relies on transitions between two neighboring rungs of the Jaynes-Cummings ladder (typically the $n_\pm = \left\{0,1\right\}$ subspaces). The appropriate drive frequencies are given by
\begin{equation}
\begin{array}{rcl}
\omega^d_{\ket{T_0}}(\pm) & = & \frac{1}{2} \left\{ \omega^c_\pm + \left[ \tilde{\omega}^q + 2\chi_\pm \right] -\delta \right\} \\
\omega^d_{\ket{S}}(\pm) & = & \frac{1}{2} \left\{ \omega^c_\pm + \left[ \tilde{\omega}^q + 2\chi_\pm \right] +\delta \right\} \\
\end{array}\
\label{eq:freqs}
\end{equation}
where $\chi_\pm$ is a cross-Kerr term leading to a $n_\pm$-dependent shift in the effective qubit frequency and $\tilde{\omega}^q$ represents the dressed qubit frequency, which has a power-dependent red shift due to the off-resonant displacement of the cavity field by the drive \footnote{Because the qubit-cavity couplings $g_i$ differ, the qubit frequencies shift by different amounts when exposed to the same intra-cavity field.  To correct for this, we place the bare qubit frequencies slightly off of resonance such that the \textit{dressed} qubit frequencies $\tilde{\omega}^q_i$ are identical.  This adjustment is power-dependent, but is on the order of 1 MHz.}. 
When a microwave drive is applied at one of these frequencies, a two-photon transition is created between the un-driven ground state $\ket{0,0, T_-}$ and the resonant partner state $\ket{\psi} \in \{\ket{1,0,S}, \ket{1,0,T_0}, \ket{0,1,S}, \ket{0,1,T_0}\}$. However, when $n_\pm > 0$ the cavities decay stochastically and irreversibly at a rate $\kappa_+ (\kappa_-) = 2\pi \times$ 650 (820) kHz to $\ket{0,0, T_0}$ or $\ket{0,0,S}$, where the final qubit state is the same as that of the state targeted by the pump.  There are no transitions from this state that are resonant with the drive. In the case of a $T_1$ decay, the drive rapidly repumps the qubits, thus creating a stabilized entangled state.  A weak off-resonant pumping into $\ket{T_+}$, which is depleted by $T_1$ rather than by active cooling, sets an upper limit on the cooling rate.  

\begin{figure}
\centering
\includegraphics[width = 8.6cm] {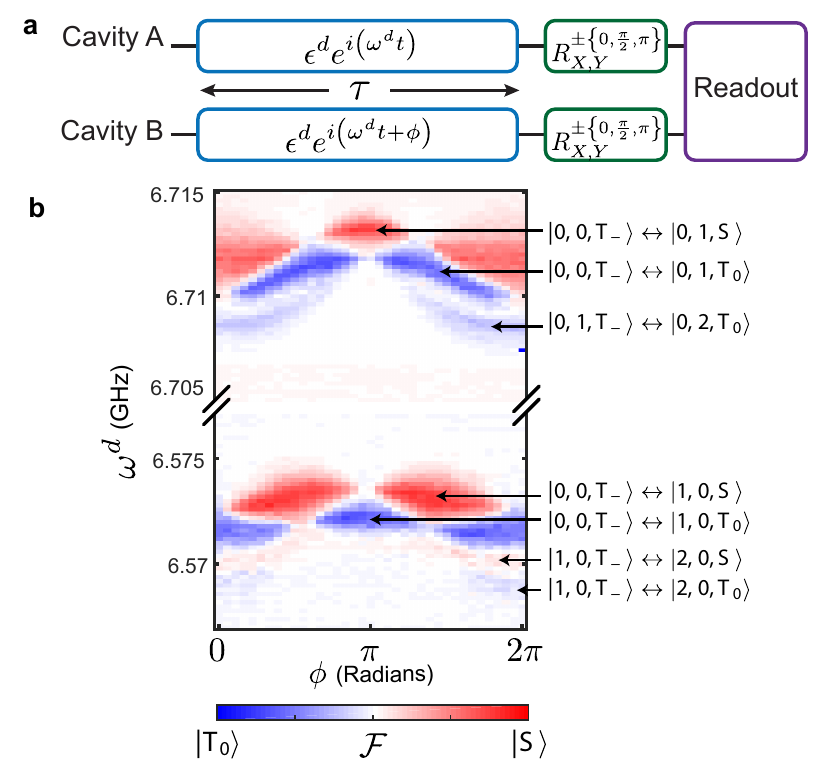}
\caption{Symmetry- and frequency- selective bath engineering.  (a) The sequence of drive, qubit, and cavity pulses used in the experiment.  We apply the bath drive for time $\tau$, then perform one of a set of tomographic rotations followed by a projective readout.  (b) Symmetry and frequency dependence of the cooling drive.  We plot $\mathcal{F}_{\ket{S}}-\mathcal{F}_{\ket{T_0}}$ such that $\ket{S}$ is red and $\ket{T_0}$ is blue. At the symmetry points $\phi=0, \phi=\pi$, and $\overline{n}_+ = \overline{n}_-$ \cite{\supp}, the drive is both frequency- and symmetry- selective.  The $\ket{\psi_1} \leftrightarrow \ket{\psi_2}$ notation indicates the transition with which the drive is resonant for the labelled band.  Transitions between higher cavity occupation states are red-detuned by $\chi_+ = 2\pi \times$ 2.5 MHz for the lower-frequency bands, and $\chi_- = 2\pi \times$ 1.4 MHz for the higher-frequency bands.}
\label{fig:flower}
\end{figure}

In Figure \ref{fig:flower}, we implement this protocol by applying simultaneous, amplitude-balanced drives with a relative phase $\phi \equiv \phi_B-\phi_A$ to the input of the cavities.  Panel (a) shows the sequence of pulses: we apply the bath drive for a fixed interval of $\tau = 10$ $\mu s$, and sweep the drive frequency ($\omega^d$, $y$-axis) and relative phase ($\phi$, $x$-axis).  We then reconstruct the joint qubit density matrix $\rho$ using tomographic reconstruction techniques \cite{Chow2010, Steffen2006} based on high-power readout \cite{Reed2010}. Figure \ref{fig:flower}(b) shows the fidelity to $\ket{S}$ (red) and to $\ket{T_0}$ (blue), where the fidelity to a target state $\ket{\psi}$ is given by $\mathcal{F} = \langle \psi| \rho | \psi \rangle$.  

The \textit{symmetry-selective} aspect of the protocol manifests at three symmetry points. In particular, there are four bands in which the protocol achieves entanglement, corresponding to the frequencies in Equation \ref{eq:freqs}: entanglement via $\omega^c_+$ $(\omega^c_-)$ occurs at $\omega^d = 2\pi \times 6.572 \text{ }(6.713) \pm 0.0013$ GHz.  However, at $\phi = 0, \phi = \pi$, and $\phi \sim 180^\circ \pm 67^\circ$, the resonant transitions are selectively suppressed for  one of the target states, and the suppressed states are reversed between the $\omega^c_+$ and $\omega^c_-$ cooling bands.  

The phase-dependent suppression can be understood as a parity selection rule that is dynamically generated by altering the drive profile across the coupled cavities.  The starting permutation-exchange parity is comprised of the initial qubit state ($\ket{T_-}$, a symmetric state) and the two photons used to generate the drive (which vary from symmetric to antisymmetric with $\phi$); the output parity is comprised of the qubit state symmetry and the dissipated photon.  Conservation of parity requires that the net parity of the output state respect that of the input state - remembering that the net exchange symmetry of two antisymmetric components is overall even.  By varying the relative phase of the drives, we vary the input symmetry and therefore control the parity selection rules.

Under an even-parity drive, when the cooling drive is comprised of two symmetric or two antisymmetric photons (\textit{i.e.} $\phi = 0$ or $\pi$), we can only cool to the qubit state whose parity is the same as the cavity output photon. Indeed, population in the antisymmetric $\ket{S}$ is fully suppressed in the lower (symmetric) band at $\phi = \{0,\pi\}$, and $\ket{T_0}$ is similarly suppressed in the upper band (where the scattered photon is antisymmetric). There also exits a relative phase at which the drive is comprised equally of symmetric and antisymmetric photons, leading to an overall odd-parity drive. This phase is $\phi \approx 180^\circ \pm 67^\circ$ in these experiments, and differs from $\pi \pm \pi/2$ because of the detuning $\omega^c_- \neq \omega^c_+$ \cite{\supp}.  At these phases, the parity of the target qubit state must be \textit{opposite} that of the cavity output photon. Cooling to $\ket{T_0}$ occurs only via the anti-symmetric cavity in this case, and cooling to $\ket{S}$ occurs via the $\omega^c_+$ mode.  These symmetry restrictions are lifted for generic $\phi$, in which case both cavity modes can be equivalently used to target $\ket{T_0}$ or $\ket{S}$, and only energy conservation of input and output photons is required.  Thus, simply by tuning a readily-adjustable parameter in our driving profile, we can turn a given entangled state from a forbidden into a symmetry-protected state.

The undulation in the cooling bands is an effect of the phase-dependence of $\tilde{\omega}^q$, due to the detuning between $\omega^c_+$ and $\omega^c_-$: a drive of fixed amplitude is closer in frequency and therefore coupled more strongly to the lower-frequency symmetric mode, resulting in a stronger AC Stark shift at $\phi \approx 0$.  The broadening of the cooling spectrum at $\phi = 0$ represents the same phenomenon, this time manifesting as power-broadening.  The faint red-shifted cooling bands, detuned by $\chi_\pm$, represent cooling between higher photon-number subspaces, as labeled.

\begin{figure}
\centering
\includegraphics[width = 8.6cm] {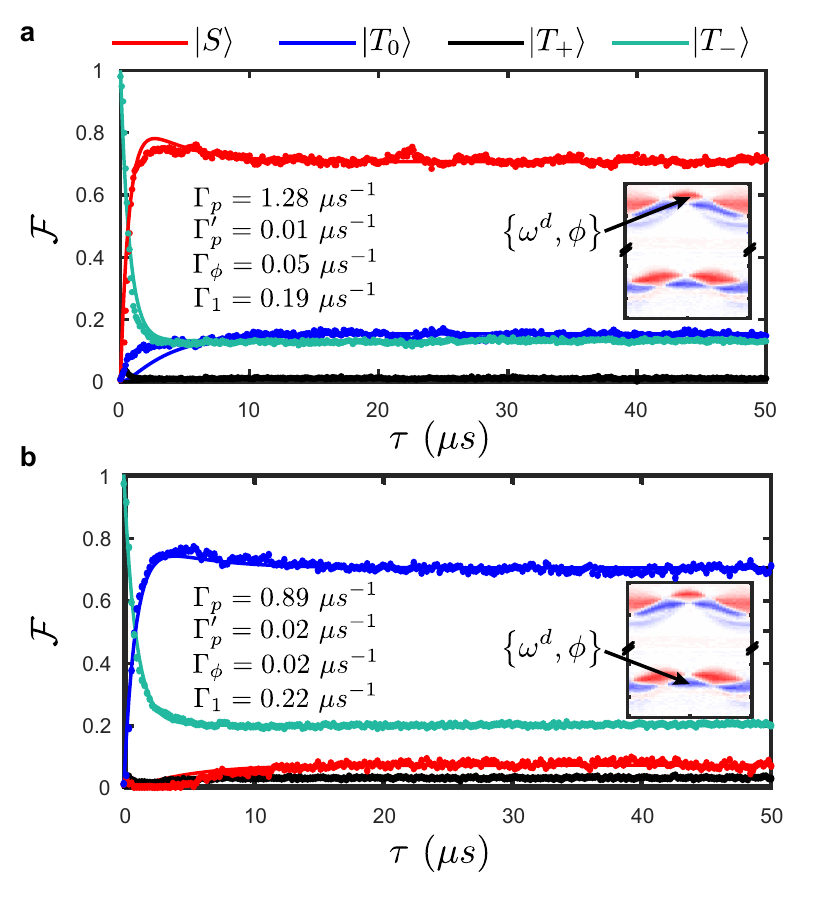}
\caption{Cooling dynamics. (a) We prepare $\ket{S}$ using $\phi=\pi$ by cooling via the antisymmetric cavity mode (inset).  (b)  Similarly, we prepare $\ket{T_0}$ using $\phi = \pi$ via the symmetric cavity mode.  In both cases, we fix $\phi$ and $\omega^d$; apply the drive for time $\tau$; and then tomographically reconstruct the resultant joint qubit state.  The experimental data are represented as dots; solid lines are fits to a coupled rate equation with rates as noted.  }
\label{fig:dynamics}
\end{figure}

By moving to the time-domain (Figure \ref{fig:dynamics}), we can resolve the effects of the several dynamical rates that govern the non-equilibrium steady state. For each experiment we fix $\omega^d$ and $\phi$, and apply the bath drive for a variable time $\tau$, again finally tomographically reconstructing the joint qubit state. We utilize $\phi = \pi$ such that parity rules require cooling to $\ket{S}$ $\left(\ket{T_0}\right)$ via $\omega^c_-$ $\left(\omega^c_+\right)$.  The dominant rates in the system are $\Gamma_p$, the pumping rate from $\ket{T_-}$ to the target state; $\Gamma_p'$, a weak off-resonant pumping to $\ket{T_+}$; $\Gamma_1$, the spontaneous decay rates of the qubits; and $\Gamma_{\phi}$, the effective dephasing rate between $\ket{S}$ and $\ket{T_0}$.  Provided that we meet the inequality $\Gamma_\phi, \Gamma_p' <\Gamma_1 <\Gamma_p$, we expect the steady-state population to be entangled. We fit the data to a coupled rate equation and extract the pumping and decay rates. The quality of the fit to a simple exponential indicates that the dynamics of this system are dominated by incoherent processes, which is consistent with $\kappa_\pm \gg \Gamma_p$: in this regime, photons stochastically leak from the cavity much more quickly than the drive is able to repopulate them. 

The steady state saturates to the entangled state of choice after a transient ring-up (dominated by $\Gamma_p$) and a small-overshoot (related to $\Gamma_\phi$). The steady-state fidelities are $\mathcal{F}(\ket{T_0}) = 0.70$ and $\mathcal{F}(\ket{S}) = 0.71$, well beyond the threshold $F=0.5$ indicative of quantum entanglement.  The fidelity loss is dominated by residual $\ket{T_-}$ population and by transitions to the entangled state of opposite symmetry $\ket{T_0} \leftrightarrow \ket{S}$.  Increasing $\Gamma_p$ in principle helps to depopulate the inital $\ket{T_-}$ state; however, increasing the pump leads to power-broadening of both the desired transition and of the off-resonant pumping to $\ket{T_+}$.  Since $\ket{T_+}$ decays equally to $\ket{S}$ and to $\ket{T_0}$, this creates a drive-dependent dephasing that sets an upper limit on the pumping rate. 
In a future on-chip implementation with currently-accessible qubit coherence times, this protocol can be expected to produce on-demand entanglement with fidelity in excess of 0.90.

In this work, we have demonstrated symmetry-selective bath engineering, harnessing both the spatial symmetry and the density of states of the dissipative environment to achieve and preserve on-demand entanglement. The engineered symmetries in our system distinguish it from the two-qubit bath engineering experiment in Ref. \cite{Shankar2013}, where cooling to $\ket{S}$ is achieved by utilizing far-detuned qubits in a single cavity; stabilizing entanglement in this system required six microwave drives, and only $\ket{S}$ was accessible.  In our implementation, the resonant construction of the photonic lattice imprints itself onto the effective qubit Hamiltonian and lifts the degeneracy in the single-excitation subspace.  The lifting of this degeneracy allows us to reduce the number of required drives from six to one, and the use of separate cavities allows us to easily modify the spatial profile of this drive in order to capitalize on the permutation symmetries of the coupled cavity resonances.  

Our work demonstrates that engineering symmetries of a dissipative environment provides a powerful route to quantum control.  Because we achieve entanglement in a resource-efficient manner, our protocol is amenable to scaling to multiple qubits and cavities.  In a multi-qubit implementation, the even and odd permutation symmetries generalize to a quasi-momentum across the lattice - and critically, adjusting the profile of a driving tone across the lattice still provides symmetry selectivity.  Furthermore, this architecture is well-suited to the preparation of many-body $W$-states \cite{Aron2014b}, the protection of high-symmetry states (i.e. quadrupoles), and the study of driven-dissipative dynamics in quantum lattices.  The ease of access to single-qubit manipulation and readout makes this experimental geometry a promising testbed for transport and studies of long-range entanglement in these systems.

\begin{acknowledgments}
We acknowledge useful discussions with Vinay Ramasesh and Shyam Shankar. This research is based on work supported in part by the U.S. Army Research Office (ARO) under grant no. W911NF-14-1-0078.  MES acknowledges support from the Fannie and John Hertz Foundation; LM acknowledges support from the Berkeley Fellowship and the National Science Foundation (NSF) Graduate Research Fellowship. HET acknowledges support from ARO Grant No. W911NF-15-1-0299 and NSF Grant No. DMR-1151810. CA acknowledges support from NSF Grant No. DMR-1151810 and the Eric and Wendy Schmidt Transformative Technology Fund.  MK gratefully acknowledges support from the Professional Staff Congress of the City University of New York award No. 68193-0046.
\end{acknowledgments}

\bibliography{parityBEbiblio}

\begin{thebibliography}{29}%
\makeatletter
\providecommand \@ifxundefined [1]{%
 \@ifx{#1\undefined}
}%
\providecommand \@ifnum [1]{%
 \ifnum #1\expandafter \@firstoftwo
 \else \expandafter \@secondoftwo
 \fi
}%
\providecommand \@ifx [1]{%
 \ifx #1\expandafter \@firstoftwo
 \else \expandafter \@secondoftwo
 \fi
}%
\providecommand \natexlab [1]{#1}%
\providecommand \enquote  [1]{``#1''}%
\providecommand \bibnamefont  [1]{#1}%
\providecommand \bibfnamefont [1]{#1}%
\providecommand \citenamefont [1]{#1}%
\providecommand \href@noop [0]{\@secondoftwo}%
\providecommand \href [0]{\begingroup \@sanitize@url \@href}%
\providecommand \@href[1]{\@@startlink{#1}\@@href}%
\providecommand \@@href[1]{\endgroup#1\@@endlink}%
\providecommand \@sanitize@url [0]{\catcode `\\12\catcode `\$12\catcode
  `\&12\catcode `\#12\catcode `\^12\catcode `\_12\catcode `\%12\relax}%
\providecommand \@@startlink[1]{}%
\providecommand \@@endlink[0]{}%
\providecommand \url  [0]{\begingroup\@sanitize@url \@url }%
\providecommand \@url [1]{\endgroup\@href {#1}{\urlprefix }}%
\providecommand \urlprefix  [0]{URL }%
\providecommand \Eprint [0]{\href }%
\providecommand \doibase [0]{http://dx.doi.org/}%
\providecommand \selectlanguage [0]{\@gobble}%
\providecommand \bibinfo  [0]{\@secondoftwo}%
\providecommand \bibfield  [0]{\@secondoftwo}%
\providecommand \translation [1]{[#1]}%
\providecommand \BibitemOpen [0]{}%
\providecommand \bibitemStop [0]{}%
\providecommand \bibitemNoStop [0]{.\EOS\space}%
\providecommand \EOS [0]{\spacefactor3000\relax}%
\providecommand \BibitemShut  [1]{\csname bibitem#1\endcsname}%
\let\auto@bib@innerbib\@empty
\bibitem [{\citenamefont {Koch}\ \emph {et~al.}(2007)\citenamefont {Koch},
  \citenamefont {Yu}, \citenamefont {Gambetta}, \citenamefont {Houck},
  \citenamefont {Schuster}, \citenamefont {Majer}, \citenamefont {Blais},
  \citenamefont {Devoret}, \citenamefont {Girvin},\ and\ \citenamefont
  {Schoelkopf}}]{Koch2007}%
  \BibitemOpen
  \bibfield  {author} {\bibinfo {author} {\bibfnamefont {J.}~\bibnamefont
  {Koch}}, \bibinfo {author} {\bibfnamefont {T.~M.}\ \bibnamefont {Yu}},
  \bibinfo {author} {\bibfnamefont {J.}~\bibnamefont {Gambetta}}, \bibinfo
  {author} {\bibfnamefont {A.~A.}\ \bibnamefont {Houck}}, \bibinfo {author}
  {\bibfnamefont {D.~I.}\ \bibnamefont {Schuster}}, \bibinfo {author}
  {\bibfnamefont {J.}~\bibnamefont {Majer}}, \bibinfo {author} {\bibfnamefont
  {A.}~\bibnamefont {Blais}}, \bibinfo {author} {\bibfnamefont {M.~H.}\
  \bibnamefont {Devoret}}, \bibinfo {author} {\bibfnamefont {S.~M.}\
  \bibnamefont {Girvin}}, \ and\ \bibinfo {author} {\bibfnamefont {R.~J.}\
  \bibnamefont {Schoelkopf}},\ }\href {\doibase 10.1103/PhysRevA.76.042319}
  {\bibfield  {journal} {\bibinfo  {journal} {Physical Review A}\ }\textbf
  {\bibinfo {volume} {76}},\ \bibinfo {pages} {042319} (\bibinfo {year}
  {2007})}\BibitemShut {NoStop}%
\bibitem [{\citenamefont {Manucharyan}\ \emph {et~al.}(2009)\citenamefont
  {Manucharyan}, \citenamefont {Koch}, \citenamefont {Glazman},\ and\
  \citenamefont {Devoret}}]{Manucharyan2009}%
  \BibitemOpen
  \bibfield  {author} {\bibinfo {author} {\bibfnamefont {V.~E.}\ \bibnamefont
  {Manucharyan}}, \bibinfo {author} {\bibfnamefont {J.}~\bibnamefont {Koch}},
  \bibinfo {author} {\bibfnamefont {L.~I.}\ \bibnamefont {Glazman}}, \ and\
  \bibinfo {author} {\bibfnamefont {M.~H.}\ \bibnamefont {Devoret}},\ }\href
  {\doibase 10.1126/science.1175552} {\bibfield  {journal} {\bibinfo  {journal}
  {Science (New York, N.Y.)}\ }\textbf {\bibinfo {volume} {326}},\ \bibinfo
  {pages} {113} (\bibinfo {year} {2009})}\BibitemShut {NoStop}%
\bibitem [{\citenamefont {Paik}\ \emph {et~al.}(2011)\citenamefont {Paik},
  \citenamefont {Schuster}, \citenamefont {Bishop}, \citenamefont {Kirchmair},
  \citenamefont {Catelani}, \citenamefont {Sears}, \citenamefont {Johnson},
  \citenamefont {Reagor}, \citenamefont {Frunzio}, \citenamefont {Glazman},
  \citenamefont {Girvin}, \citenamefont {Devoret},\ and\ \citenamefont
  {Schoelkopf}}]{Paik2011}%
  \BibitemOpen
  \bibfield  {author} {\bibinfo {author} {\bibfnamefont {H.}~\bibnamefont
  {Paik}}, \bibinfo {author} {\bibfnamefont {D.~I.}\ \bibnamefont {Schuster}},
  \bibinfo {author} {\bibfnamefont {L.~S.}\ \bibnamefont {Bishop}}, \bibinfo
  {author} {\bibfnamefont {G.}~\bibnamefont {Kirchmair}}, \bibinfo {author}
  {\bibfnamefont {G.}~\bibnamefont {Catelani}}, \bibinfo {author}
  {\bibfnamefont {A.~P.}\ \bibnamefont {Sears}}, \bibinfo {author}
  {\bibfnamefont {B.~R.}\ \bibnamefont {Johnson}}, \bibinfo {author}
  {\bibfnamefont {M.~J.}\ \bibnamefont {Reagor}}, \bibinfo {author}
  {\bibfnamefont {L.}~\bibnamefont {Frunzio}}, \bibinfo {author} {\bibfnamefont
  {L.~I.}\ \bibnamefont {Glazman}}, \bibinfo {author} {\bibfnamefont {S.~M.}\
  \bibnamefont {Girvin}}, \bibinfo {author} {\bibfnamefont {M.~H.}\
  \bibnamefont {Devoret}}, \ and\ \bibinfo {author} {\bibfnamefont {R.~J.}\
  \bibnamefont {Schoelkopf}},\ }\href {\doibase 10.1103/PhysRevLett.107.240501}
  {\bibfield  {journal} {\bibinfo  {journal} {Physical Review Letters}\
  }\textbf {\bibinfo {volume} {107}},\ \bibinfo {pages} {240501} (\bibinfo
  {year} {2011})}\BibitemShut {NoStop}%
\bibitem [{\citenamefont {Barends}\ \emph {et~al.}(2013)\citenamefont
  {Barends}, \citenamefont {Kelly}, \citenamefont {Megrant}, \citenamefont
  {Sank}, \citenamefont {Jeffrey}, \citenamefont {Chen}, \citenamefont {Yin},
  \citenamefont {Chiaro}, \citenamefont {Mutus}, \citenamefont {Neill},
  \citenamefont {O’Malley}, \citenamefont {Roushan}, \citenamefont {Wenner},
  \citenamefont {White}, \citenamefont {Cleland},\ and\ \citenamefont
  {Martinis}}]{Barends2013}%
  \BibitemOpen
  \bibfield  {author} {\bibinfo {author} {\bibfnamefont {R.}~\bibnamefont
  {Barends}}, \bibinfo {author} {\bibfnamefont {J.}~\bibnamefont {Kelly}},
  \bibinfo {author} {\bibfnamefont {A.}~\bibnamefont {Megrant}}, \bibinfo
  {author} {\bibfnamefont {D.}~\bibnamefont {Sank}}, \bibinfo {author}
  {\bibfnamefont {E.}~\bibnamefont {Jeffrey}}, \bibinfo {author} {\bibfnamefont
  {Y.}~\bibnamefont {Chen}}, \bibinfo {author} {\bibfnamefont {Y.}~\bibnamefont
  {Yin}}, \bibinfo {author} {\bibfnamefont {B.}~\bibnamefont {Chiaro}},
  \bibinfo {author} {\bibfnamefont {J.}~\bibnamefont {Mutus}}, \bibinfo
  {author} {\bibfnamefont {C.}~\bibnamefont {Neill}}, \bibinfo {author}
  {\bibfnamefont {P.}~\bibnamefont {O’Malley}}, \bibinfo {author}
  {\bibfnamefont {P.}~\bibnamefont {Roushan}}, \bibinfo {author} {\bibfnamefont
  {J.}~\bibnamefont {Wenner}}, \bibinfo {author} {\bibfnamefont {T.~C.}\
  \bibnamefont {White}}, \bibinfo {author} {\bibfnamefont {A.~N.}\ \bibnamefont
  {Cleland}}, \ and\ \bibinfo {author} {\bibfnamefont {J.~M.}\ \bibnamefont
  {Martinis}},\ }\href {\doibase 10.1103/PhysRevLett.111.080502} {\bibfield
  {journal} {\bibinfo  {journal} {Physical Review Letters}\ }\textbf {\bibinfo
  {volume} {111}},\ \bibinfo {pages} {080502} (\bibinfo {year}
  {2013})}\BibitemShut {NoStop}%
\bibitem [{\citenamefont {Barends}\ \emph {et~al.}(2014)\citenamefont
  {Barends}, \citenamefont {Kelly}, \citenamefont {Megrant}, \citenamefont
  {Veitia}, \citenamefont {Sank}, \citenamefont {Jeffrey}, \citenamefont
  {White}, \citenamefont {Mutus}, \citenamefont {Fowler}, \citenamefont
  {Campbell}, \citenamefont {Chen}, \citenamefont {Chen}, \citenamefont
  {Chiaro}, \citenamefont {Dunsworth}, \citenamefont {Neill}, \citenamefont
  {O'Malley}, \citenamefont {Roushan}, \citenamefont {Vainsencher},
  \citenamefont {Wenner}, \citenamefont {Korotkov}, \citenamefont {Cleland},\
  and\ \citenamefont {Martinis}}]{Barends2014}%
  \BibitemOpen
  \bibfield  {author} {\bibinfo {author} {\bibfnamefont {R.}~\bibnamefont
  {Barends}}, \bibinfo {author} {\bibfnamefont {J.}~\bibnamefont {Kelly}},
  \bibinfo {author} {\bibfnamefont {A.}~\bibnamefont {Megrant}}, \bibinfo
  {author} {\bibfnamefont {A.}~\bibnamefont {Veitia}}, \bibinfo {author}
  {\bibfnamefont {D.}~\bibnamefont {Sank}}, \bibinfo {author} {\bibfnamefont
  {E.}~\bibnamefont {Jeffrey}}, \bibinfo {author} {\bibfnamefont {T.~C.}\
  \bibnamefont {White}}, \bibinfo {author} {\bibfnamefont {J.}~\bibnamefont
  {Mutus}}, \bibinfo {author} {\bibfnamefont {A.~G.}\ \bibnamefont {Fowler}},
  \bibinfo {author} {\bibfnamefont {B.}~\bibnamefont {Campbell}}, \bibinfo
  {author} {\bibfnamefont {Y.}~\bibnamefont {Chen}}, \bibinfo {author}
  {\bibfnamefont {Z.}~\bibnamefont {Chen}}, \bibinfo {author} {\bibfnamefont
  {B.}~\bibnamefont {Chiaro}}, \bibinfo {author} {\bibfnamefont
  {A.}~\bibnamefont {Dunsworth}}, \bibinfo {author} {\bibfnamefont
  {C.}~\bibnamefont {Neill}}, \bibinfo {author} {\bibfnamefont
  {P.}~\bibnamefont {O'Malley}}, \bibinfo {author} {\bibfnamefont
  {P.}~\bibnamefont {Roushan}}, \bibinfo {author} {\bibfnamefont
  {A.}~\bibnamefont {Vainsencher}}, \bibinfo {author} {\bibfnamefont
  {J.}~\bibnamefont {Wenner}}, \bibinfo {author} {\bibfnamefont {A.~N.}\
  \bibnamefont {Korotkov}}, \bibinfo {author} {\bibfnamefont {A.~N.}\
  \bibnamefont {Cleland}}, \ and\ \bibinfo {author} {\bibfnamefont {J.~M.}\
  \bibnamefont {Martinis}},\ }\href {\doibase 10.1038/nature13171} {\bibfield
  {journal} {\bibinfo  {journal} {Nature}\ }\textbf {\bibinfo {volume} {508}},\
  \bibinfo {pages} {500} (\bibinfo {year} {2014})}\BibitemShut {NoStop}%
\bibitem [{\citenamefont {C\'{o}rcoles}\ \emph {et~al.}(2015)\citenamefont
  {C\'{o}rcoles}, \citenamefont {Magesan}, \citenamefont {Srinivasan},
  \citenamefont {Cross}, \citenamefont {Steffen}, \citenamefont {Gambetta},\
  and\ \citenamefont {Chow}}]{Corcoles2015}%
  \BibitemOpen
  \bibfield  {author} {\bibinfo {author} {\bibfnamefont {A.}~\bibnamefont
  {C\'{o}rcoles}}, \bibinfo {author} {\bibfnamefont {E.}~\bibnamefont
  {Magesan}}, \bibinfo {author} {\bibfnamefont {S.~J.}\ \bibnamefont
  {Srinivasan}}, \bibinfo {author} {\bibfnamefont {A.~W.}\ \bibnamefont
  {Cross}}, \bibinfo {author} {\bibfnamefont {M.}~\bibnamefont {Steffen}},
  \bibinfo {author} {\bibfnamefont {J.~M.}\ \bibnamefont {Gambetta}}, \ and\
  \bibinfo {author} {\bibfnamefont {J.~M.}\ \bibnamefont {Chow}},\ }\href
  {\doibase 10.1038/ncomms7979} {\bibfield  {journal} {\bibinfo  {journal}
  {Nature Communications}\ }\textbf {\bibinfo {volume} {6}},\ \bibinfo {pages}
  {6979} (\bibinfo {year} {2015})}\BibitemShut {NoStop}%
\bibitem [{\citenamefont {Rist\`{e}}\ \emph {et~al.}(2015)\citenamefont
  {Rist\`{e}}, \citenamefont {Poletto}, \citenamefont {Huang}, \citenamefont
  {Bruno}, \citenamefont {Vesterinen}, \citenamefont {Saira},\ and\
  \citenamefont {DiCarlo}}]{Riste2015}%
  \BibitemOpen
  \bibfield  {author} {\bibinfo {author} {\bibfnamefont {D.}~\bibnamefont
  {Rist\`{e}}}, \bibinfo {author} {\bibfnamefont {S.}~\bibnamefont {Poletto}},
  \bibinfo {author} {\bibfnamefont {M.-Z.}\ \bibnamefont {Huang}}, \bibinfo
  {author} {\bibfnamefont {A.}~\bibnamefont {Bruno}}, \bibinfo {author}
  {\bibfnamefont {V.}~\bibnamefont {Vesterinen}}, \bibinfo {author}
  {\bibfnamefont {O.-P.}\ \bibnamefont {Saira}}, \ and\ \bibinfo {author}
  {\bibfnamefont {L.}~\bibnamefont {DiCarlo}},\ }\href {\doibase
  10.1038/ncomms7983} {\bibfield  {journal} {\bibinfo  {journal} {Nature
  Communications}\ }\textbf {\bibinfo {volume} {6}},\ \bibinfo {pages} {6983}
  (\bibinfo {year} {2015})}\BibitemShut {NoStop}%
\bibitem [{\citenamefont {Poyatos}\ \emph {et~al.}(1996)\citenamefont
  {Poyatos}, \citenamefont {Cirac},\ and\ \citenamefont
  {Zoller}}]{Poyatos1996}%
  \BibitemOpen
  \bibfield  {author} {\bibinfo {author} {\bibfnamefont {J.~F.}\ \bibnamefont
  {Poyatos}}, \bibinfo {author} {\bibfnamefont {J.~I.}\ \bibnamefont {Cirac}},
  \ and\ \bibinfo {author} {\bibfnamefont {P.}~\bibnamefont {Zoller}},\ }\href
  {\doibase 10.1103/PhysRevLett.77.4728} {\bibfield  {journal} {\bibinfo
  {journal} {Physical Review Letters}\ }\textbf {\bibinfo {volume} {77}},\
  \bibinfo {pages} {4728} (\bibinfo {year} {1996})}\BibitemShut {NoStop}%
\bibitem [{\citenamefont {Lin}\ \emph {et~al.}(2013)\citenamefont {Lin},
  \citenamefont {Gaebler}, \citenamefont {Reiter}, \citenamefont {Tan},
  \citenamefont {Bowler}, \citenamefont {S\o~rensen}, \citenamefont
  {Leibfried},\ and\ \citenamefont {Wineland}}]{Lin2013}%
  \BibitemOpen
  \bibfield  {author} {\bibinfo {author} {\bibfnamefont {Y.}~\bibnamefont
  {Lin}}, \bibinfo {author} {\bibfnamefont {J.~P.}\ \bibnamefont {Gaebler}},
  \bibinfo {author} {\bibfnamefont {F.}~\bibnamefont {Reiter}}, \bibinfo
  {author} {\bibfnamefont {T.~R.}\ \bibnamefont {Tan}}, \bibinfo {author}
  {\bibfnamefont {R.}~\bibnamefont {Bowler}}, \bibinfo {author} {\bibfnamefont
  {A.~S.}\ \bibnamefont {S\o~rensen}}, \bibinfo {author} {\bibfnamefont
  {D.}~\bibnamefont {Leibfried}}, \ and\ \bibinfo {author} {\bibfnamefont
  {D.~J.}\ \bibnamefont {Wineland}},\ }\href {\doibase 10.1038/nature12801}
  {\bibfield  {journal} {\bibinfo  {journal} {Nature}\ }\textbf {\bibinfo
  {volume} {504}},\ \bibinfo {pages} {415} (\bibinfo {year}
  {2013})}\BibitemShut {NoStop}%
\bibitem [{\citenamefont {Krauter}\ \emph {et~al.}(2011)\citenamefont
  {Krauter}, \citenamefont {Muschik}, \citenamefont {Jensen}, \citenamefont
  {Wasilewski}, \citenamefont {Petersen}, \citenamefont {Cirac},\ and\
  \citenamefont {Polzik}}]{Krauter2011}%
  \BibitemOpen
  \bibfield  {author} {\bibinfo {author} {\bibfnamefont {H.}~\bibnamefont
  {Krauter}}, \bibinfo {author} {\bibfnamefont {C.~A.}\ \bibnamefont
  {Muschik}}, \bibinfo {author} {\bibfnamefont {K.}~\bibnamefont {Jensen}},
  \bibinfo {author} {\bibfnamefont {W.}~\bibnamefont {Wasilewski}}, \bibinfo
  {author} {\bibfnamefont {J.~M.}\ \bibnamefont {Petersen}}, \bibinfo {author}
  {\bibfnamefont {J.~I.}\ \bibnamefont {Cirac}}, \ and\ \bibinfo {author}
  {\bibfnamefont {E.~S.}\ \bibnamefont {Polzik}},\ }\href {\doibase
  10.1103/PhysRevLett.107.080503} {\bibfield  {journal} {\bibinfo  {journal}
  {Physical Review Letters}\ }\textbf {\bibinfo {volume} {107}},\ \bibinfo
  {pages} {080503} (\bibinfo {year} {2011})}\BibitemShut {NoStop}%
\bibitem [{\citenamefont {Barreiro}\ \emph {et~al.}(2011)\citenamefont
  {Barreiro}, \citenamefont {M{\"{u}}ller}, \citenamefont {Schindler},
  \citenamefont {Nigg}, \citenamefont {Monz}, \citenamefont {Chwalla},
  \citenamefont {Hennrich}, \citenamefont {Roos}, \citenamefont {Zoller},\ and\
  \citenamefont {Blatt}}]{Barreiro2011}%
  \BibitemOpen
  \bibfield  {author} {\bibinfo {author} {\bibfnamefont {J.~T.}\ \bibnamefont
  {Barreiro}}, \bibinfo {author} {\bibfnamefont {M.}~\bibnamefont
  {M{\"{u}}ller}}, \bibinfo {author} {\bibfnamefont {P.}~\bibnamefont
  {Schindler}}, \bibinfo {author} {\bibfnamefont {D.}~\bibnamefont {Nigg}},
  \bibinfo {author} {\bibfnamefont {T.}~\bibnamefont {Monz}}, \bibinfo {author}
  {\bibfnamefont {M.}~\bibnamefont {Chwalla}}, \bibinfo {author} {\bibfnamefont
  {M.}~\bibnamefont {Hennrich}}, \bibinfo {author} {\bibfnamefont {C.~F.}\
  \bibnamefont {Roos}}, \bibinfo {author} {\bibfnamefont {P.}~\bibnamefont
  {Zoller}}, \ and\ \bibinfo {author} {\bibfnamefont {R.}~\bibnamefont
  {Blatt}},\ }\href {\doibase 10.1038/nature09801} {\bibfield  {journal}
  {\bibinfo  {journal} {Nature}\ }\textbf {\bibinfo {volume} {470}},\ \bibinfo
  {pages} {486} (\bibinfo {year} {2011})}\BibitemShut {NoStop}%
\bibitem [{\citenamefont {Murch}\ \emph {et~al.}(2012)\citenamefont {Murch},
  \citenamefont {Vool}, \citenamefont {Zhou}, \citenamefont {Weber},
  \citenamefont {Girvin},\ and\ \citenamefont {Siddiqi}}]{Murch2012}%
  \BibitemOpen
  \bibfield  {author} {\bibinfo {author} {\bibfnamefont {K.~W.}\ \bibnamefont
  {Murch}}, \bibinfo {author} {\bibfnamefont {U.}~\bibnamefont {Vool}},
  \bibinfo {author} {\bibfnamefont {D.}~\bibnamefont {Zhou}}, \bibinfo {author}
  {\bibfnamefont {S.~J.}\ \bibnamefont {Weber}}, \bibinfo {author}
  {\bibfnamefont {S.~M.}\ \bibnamefont {Girvin}}, \ and\ \bibinfo {author}
  {\bibfnamefont {I.}~\bibnamefont {Siddiqi}},\ }\href {\doibase
  10.1103/PhysRevLett.109.183602} {\bibfield  {journal} {\bibinfo  {journal}
  {Physical Review Letters}\ }\textbf {\bibinfo {volume} {109}},\ \bibinfo
  {pages} {183602} (\bibinfo {year} {2012})}\BibitemShut {NoStop}%
\bibitem [{\citenamefont {Shankar}\ \emph {et~al.}(2013)\citenamefont
  {Shankar}, \citenamefont {Hatridge}, \citenamefont {Leghtas}, \citenamefont
  {Sliwa}, \citenamefont {Narla}, \citenamefont {Vool}, \citenamefont {Girvin},
  \citenamefont {Frunzio}, \citenamefont {Mirrahimi},\ and\ \citenamefont
  {Devoret}}]{Shankar2013}%
  \BibitemOpen
  \bibfield  {author} {\bibinfo {author} {\bibfnamefont {S.}~\bibnamefont
  {Shankar}}, \bibinfo {author} {\bibfnamefont {M.}~\bibnamefont {Hatridge}},
  \bibinfo {author} {\bibfnamefont {Z.}~\bibnamefont {Leghtas}}, \bibinfo
  {author} {\bibfnamefont {K.~M.}\ \bibnamefont {Sliwa}}, \bibinfo {author}
  {\bibfnamefont {A.}~\bibnamefont {Narla}}, \bibinfo {author} {\bibfnamefont
  {U.}~\bibnamefont {Vool}}, \bibinfo {author} {\bibfnamefont {S.~M.}\
  \bibnamefont {Girvin}}, \bibinfo {author} {\bibfnamefont {L.}~\bibnamefont
  {Frunzio}}, \bibinfo {author} {\bibfnamefont {M.}~\bibnamefont {Mirrahimi}},
  \ and\ \bibinfo {author} {\bibfnamefont {M.~H.}\ \bibnamefont {Devoret}},\
  }\href {\doibase 10.1038/nature12802} {\bibfield  {journal} {\bibinfo
  {journal} {Nature}\ }\textbf {\bibinfo {volume} {504}},\ \bibinfo {pages}
  {419} (\bibinfo {year} {2013})}\BibitemShut {NoStop}%
\bibitem [{\citenamefont {Hacohen-Gourgy}\ \emph {et~al.}(2015)\citenamefont
  {Hacohen-Gourgy}, \citenamefont {Ramasesh}, \citenamefont {{De Grandi}},
  \citenamefont {Siddiqi},\ and\ \citenamefont {Girvin}}]{Hacohen-Gourgy2015}%
  \BibitemOpen
  \bibfield  {author} {\bibinfo {author} {\bibfnamefont {S.}~\bibnamefont
  {Hacohen-Gourgy}}, \bibinfo {author} {\bibfnamefont {V.}~\bibnamefont
  {Ramasesh}}, \bibinfo {author} {\bibfnamefont {C.}~\bibnamefont {{De
  Grandi}}}, \bibinfo {author} {\bibfnamefont {I.}~\bibnamefont {Siddiqi}}, \
  and\ \bibinfo {author} {\bibfnamefont {S.~M.}\ \bibnamefont {Girvin}},\
  }\href {http://arxiv.org/abs/1506.05837} {\ ,\ \bibinfo {pages} {5} (\bibinfo
  {year} {2015})},\ \Eprint {http://arxiv.org/abs/1506.05837}
  {arXiv:1506.05837} \BibitemShut {NoStop}%
\bibitem [{\citenamefont {Leghtas}\ \emph {et~al.}(2015)\citenamefont
  {Leghtas}, \citenamefont {Touzard}, \citenamefont {Pop}, \citenamefont {Kou},
  \citenamefont {Vlastakis}, \citenamefont {Petrenko}, \citenamefont {Sliwa},
  \citenamefont {Narla}, \citenamefont {Shankar}, \citenamefont {Hatridge},
  \citenamefont {Reagor}, \citenamefont {Frunzio}, \citenamefont {Schoelkopf},
  \citenamefont {Mirrahimi},\ and\ \citenamefont {Devoret}}]{Leghtas2015}%
  \BibitemOpen
  \bibfield  {author} {\bibinfo {author} {\bibfnamefont {Z.}~\bibnamefont
  {Leghtas}}, \bibinfo {author} {\bibfnamefont {S.}~\bibnamefont {Touzard}},
  \bibinfo {author} {\bibfnamefont {I.~M.}\ \bibnamefont {Pop}}, \bibinfo
  {author} {\bibfnamefont {A.}~\bibnamefont {Kou}}, \bibinfo {author}
  {\bibfnamefont {B.}~\bibnamefont {Vlastakis}}, \bibinfo {author}
  {\bibfnamefont {A.}~\bibnamefont {Petrenko}}, \bibinfo {author}
  {\bibfnamefont {K.~M.}\ \bibnamefont {Sliwa}}, \bibinfo {author}
  {\bibfnamefont {A.}~\bibnamefont {Narla}}, \bibinfo {author} {\bibfnamefont
  {S.}~\bibnamefont {Shankar}}, \bibinfo {author} {\bibfnamefont {M.~J.}\
  \bibnamefont {Hatridge}}, \bibinfo {author} {\bibfnamefont {M.}~\bibnamefont
  {Reagor}}, \bibinfo {author} {\bibfnamefont {L.}~\bibnamefont {Frunzio}},
  \bibinfo {author} {\bibfnamefont {R.~J.}\ \bibnamefont {Schoelkopf}},
  \bibinfo {author} {\bibfnamefont {M.}~\bibnamefont {Mirrahimi}}, \ and\
  \bibinfo {author} {\bibfnamefont {M.~H.}\ \bibnamefont {Devoret}},\ }\href
  {\doibase 10.1126/science.aaa2085} {\bibfield  {journal} {\bibinfo  {journal}
  {Science (New York, N.Y.)}\ }\textbf {\bibinfo {volume} {347}},\ \bibinfo
  {pages} {853} (\bibinfo {year} {2015})}\BibitemShut {NoStop}%
\bibitem [{\citenamefont {Holland}\ \emph {et~al.}(2015)\citenamefont
  {Holland}, \citenamefont {Vlastakis}, \citenamefont {Heeres}, \citenamefont
  {Reagor}, \citenamefont {Vool}, \citenamefont {Leghtas}, \citenamefont
  {Frunzio}, \citenamefont {Kirchmair}, \citenamefont {Devoret}, \citenamefont
  {Mirrahimi},\ and\ \citenamefont {Schoelkopf}}]{Holland2015}%
  \BibitemOpen
  \bibfield  {author} {\bibinfo {author} {\bibfnamefont {E.~T.}\ \bibnamefont
  {Holland}}, \bibinfo {author} {\bibfnamefont {B.}~\bibnamefont {Vlastakis}},
  \bibinfo {author} {\bibfnamefont {R.~W.}\ \bibnamefont {Heeres}}, \bibinfo
  {author} {\bibfnamefont {M.~J.}\ \bibnamefont {Reagor}}, \bibinfo {author}
  {\bibfnamefont {U.}~\bibnamefont {Vool}}, \bibinfo {author} {\bibfnamefont
  {Z.}~\bibnamefont {Leghtas}}, \bibinfo {author} {\bibfnamefont
  {L.}~\bibnamefont {Frunzio}}, \bibinfo {author} {\bibfnamefont
  {G.}~\bibnamefont {Kirchmair}}, \bibinfo {author} {\bibfnamefont {M.~H.}\
  \bibnamefont {Devoret}}, \bibinfo {author} {\bibfnamefont {M.}~\bibnamefont
  {Mirrahimi}}, \ and\ \bibinfo {author} {\bibfnamefont {R.~J.}\ \bibnamefont
  {Schoelkopf}},\ }\href {http://arxiv.org/abs/1504.03382} {\ ,\ \bibinfo
  {pages} {8} (\bibinfo {year} {2015})},\ \Eprint
  {http://arxiv.org/abs/1504.03382} {arXiv:1504.03382} \BibitemShut {NoStop}%
\bibitem [{\citenamefont {Cohen}\ and\ \citenamefont
  {Mirrahimi}(2014)}]{Cohen2014}%
  \BibitemOpen
  \bibfield  {author} {\bibinfo {author} {\bibfnamefont {J.}~\bibnamefont
  {Cohen}}\ and\ \bibinfo {author} {\bibfnamefont {M.}~\bibnamefont
  {Mirrahimi}},\ }\href {\doibase 10.1103/PhysRevA.90.062344} {\bibfield
  {journal} {\bibinfo  {journal} {Physical Review A}\ }\textbf {\bibinfo
  {volume} {90}},\ \bibinfo {pages} {062344} (\bibinfo {year}
  {2014})}\BibitemShut {NoStop}%
\bibitem [{\citenamefont {Kapit}\ \emph {et~al.}(2015)\citenamefont {Kapit},
  \citenamefont {Chalker},\ and\ \citenamefont {Simon}}]{Kapit2015}%
  \BibitemOpen
  \bibfield  {author} {\bibinfo {author} {\bibfnamefont {E.}~\bibnamefont
  {Kapit}}, \bibinfo {author} {\bibfnamefont {J.~T.}\ \bibnamefont {Chalker}},
  \ and\ \bibinfo {author} {\bibfnamefont {S.~H.}\ \bibnamefont {Simon}},\
  }\href {\doibase 10.1103/PhysRevA.91.062324} {\bibfield  {journal} {\bibinfo
  {journal} {Physical Review A}\ }\textbf {\bibinfo {volume} {91}},\ \bibinfo
  {pages} {062324} (\bibinfo {year} {2015})}\BibitemShut {NoStop}%
\bibitem [{\citenamefont {Kerckhoff}\ \emph {et~al.}(2010)\citenamefont
  {Kerckhoff}, \citenamefont {Nurdin}, \citenamefont {Pavlichin},\ and\
  \citenamefont {Mabuchi}}]{Kerckhoff2010}%
  \BibitemOpen
  \bibfield  {author} {\bibinfo {author} {\bibfnamefont {J.}~\bibnamefont
  {Kerckhoff}}, \bibinfo {author} {\bibfnamefont {H.~I.}\ \bibnamefont
  {Nurdin}}, \bibinfo {author} {\bibfnamefont {D.~S.}\ \bibnamefont
  {Pavlichin}}, \ and\ \bibinfo {author} {\bibfnamefont {H.}~\bibnamefont
  {Mabuchi}},\ }\href {\doibase 10.1103/PhysRevLett.105.040502} {\bibfield
  {journal} {\bibinfo  {journal} {Physical Review Letters}\ }\textbf {\bibinfo
  {volume} {105}},\ \bibinfo {pages} {040502} (\bibinfo {year}
  {2010})}\BibitemShut {NoStop}%
\bibitem [{\citenamefont {Verstraete}\ \emph {et~al.}(2009)\citenamefont
  {Verstraete}, \citenamefont {Wolf},\ and\ \citenamefont {{Ignacio
  Cirac}}}]{Verstraete2009}%
  \BibitemOpen
  \bibfield  {author} {\bibinfo {author} {\bibfnamefont {F.}~\bibnamefont
  {Verstraete}}, \bibinfo {author} {\bibfnamefont {M.~M.}\ \bibnamefont
  {Wolf}}, \ and\ \bibinfo {author} {\bibfnamefont {J.}~\bibnamefont {{Ignacio
  Cirac}}},\ }\href {\doibase 10.1038/nphys1342} {\bibfield  {journal}
  {\bibinfo  {journal} {Nature Physics}\ }\textbf {\bibinfo {volume} {5}},\
  \bibinfo {pages} {633} (\bibinfo {year} {2009})}\BibitemShut {NoStop}%
\bibitem [{\citenamefont {Mirrahimi}\ \emph {et~al.}(2014)\citenamefont
  {Mirrahimi}, \citenamefont {Leghtas}, \citenamefont {Albert}, \citenamefont
  {Touzard}, \citenamefont {Schoelkopf}, \citenamefont {Jiang},\ and\
  \citenamefont {Devoret}}]{Mirrahimi2014}%
  \BibitemOpen
  \bibfield  {author} {\bibinfo {author} {\bibfnamefont {M.}~\bibnamefont
  {Mirrahimi}}, \bibinfo {author} {\bibfnamefont {Z.}~\bibnamefont {Leghtas}},
  \bibinfo {author} {\bibfnamefont {V.~V.}\ \bibnamefont {Albert}}, \bibinfo
  {author} {\bibfnamefont {S.}~\bibnamefont {Touzard}}, \bibinfo {author}
  {\bibfnamefont {R.~J.}\ \bibnamefont {Schoelkopf}}, \bibinfo {author}
  {\bibfnamefont {L.}~\bibnamefont {Jiang}}, \ and\ \bibinfo {author}
  {\bibfnamefont {M.~H.}\ \bibnamefont {Devoret}},\ }\href {\doibase
  10.1088/1367-2630/16/4/045014} {\bibfield  {journal} {\bibinfo  {journal}
  {New Journal of Physics}\ }\textbf {\bibinfo {volume} {16}},\ \bibinfo
  {pages} {045014} (\bibinfo {year} {2014})}\BibitemShut {NoStop}%
\bibitem [{Note1()}]{Note1}%
  \BibitemOpen
  \bibinfo {note} {See supplementary information.}\BibitemShut {Stop}%
\bibitem [{\citenamefont {Aron}\ \emph
  {et~al.}(2014{\natexlab{a}})\citenamefont {Aron}, \citenamefont {Kulkarni},\
  and\ \citenamefont {T\"ureci}}]{Aron2014a}%
  \BibitemOpen
  \bibfield  {author} {\bibinfo {author} {\bibfnamefont {C.}~\bibnamefont
  {Aron}}, \bibinfo {author} {\bibfnamefont {M.}~\bibnamefont {Kulkarni}}, \
  and\ \bibinfo {author} {\bibfnamefont {H.~E.}\ \bibnamefont {T\"ureci}},\
  }\href {\doibase 10.1103/PhysRevA.90.062305} {\bibfield  {journal} {\bibinfo
  {journal} {Phys. Rev. A}\ }\textbf {\bibinfo {volume} {90}},\ \bibinfo
  {pages} {062305} (\bibinfo {year} {2014}{\natexlab{a}})}\BibitemShut
  {NoStop}%
\bibitem [{Note2()}]{Note2}%
  \BibitemOpen
  \bibinfo {note} {Since the target entangled states are in fact eigenstates of
  the coupled Hamiltonian, it is in principle possible to coherently pulse to
  these states. However, because the splitting is small, a coherent pulse with
  narrow enough bandwidth to drive selectively to one of these states would
  need to be several microseconds long, and therefore would be spoiled by qubit
  decay.}\BibitemShut {Stop}%
\bibitem [{Note3()}]{Note3}%
  \BibitemOpen
  \bibinfo {note} {Because the qubit-cavity couplings $g_i$ differ, the qubit
  frequencies shift by different amounts when exposed to the same intra-cavity
  field. To correct for this, we place the bare qubit frequencies slightly off
  of resonance such that the \protect \textit {dressed} qubit frequencies
  $\protect \mathaccentV {tilde}07E{\omega }^q_i$ are identical. This
  adjustment is power-dependent, but is on the order of 1 MHz.}\BibitemShut
  {Stop}%
\bibitem [{\citenamefont {Chow}\ \emph {et~al.}(2010)\citenamefont {Chow},
  \citenamefont {DiCarlo}, \citenamefont {Gambetta}, \citenamefont
  {Nunnenkamp}, \citenamefont {Bishop}, \citenamefont {Frunzio}, \citenamefont
  {Devoret}, \citenamefont {Girvin},\ and\ \citenamefont
  {Schoelkopf}}]{Chow2010}%
  \BibitemOpen
  \bibfield  {author} {\bibinfo {author} {\bibfnamefont {J.~M.}\ \bibnamefont
  {Chow}}, \bibinfo {author} {\bibfnamefont {L.}~\bibnamefont {DiCarlo}},
  \bibinfo {author} {\bibfnamefont {J.~M.}\ \bibnamefont {Gambetta}}, \bibinfo
  {author} {\bibfnamefont {A.}~\bibnamefont {Nunnenkamp}}, \bibinfo {author}
  {\bibfnamefont {L.~S.}\ \bibnamefont {Bishop}}, \bibinfo {author}
  {\bibfnamefont {L.}~\bibnamefont {Frunzio}}, \bibinfo {author} {\bibfnamefont
  {M.~H.}\ \bibnamefont {Devoret}}, \bibinfo {author} {\bibfnamefont {S.~M.}\
  \bibnamefont {Girvin}}, \ and\ \bibinfo {author} {\bibfnamefont {R.~J.}\
  \bibnamefont {Schoelkopf}},\ }\href {\doibase 10.1103/PhysRevA.81.062325}
  {\bibfield  {journal} {\bibinfo  {journal} {Phys. Rev. A}\ }\textbf {\bibinfo
  {volume} {81}},\ \bibinfo {pages} {062325} (\bibinfo {year}
  {2010})}\BibitemShut {NoStop}%
\bibitem [{\citenamefont {Steffen}\ \emph {et~al.}(2006)\citenamefont
  {Steffen}, \citenamefont {Ansmann}, \citenamefont {Bialczak}, \citenamefont
  {Katz}, \citenamefont {Lucero}, \citenamefont {McDermott}, \citenamefont
  {Neeley}, \citenamefont {Weig}, \citenamefont {Cleland},\ and\ \citenamefont
  {Martinis}}]{Steffen2006}%
  \BibitemOpen
  \bibfield  {author} {\bibinfo {author} {\bibfnamefont {M.}~\bibnamefont
  {Steffen}}, \bibinfo {author} {\bibfnamefont {M.}~\bibnamefont {Ansmann}},
  \bibinfo {author} {\bibfnamefont {R.~C.}\ \bibnamefont {Bialczak}}, \bibinfo
  {author} {\bibfnamefont {N.}~\bibnamefont {Katz}}, \bibinfo {author}
  {\bibfnamefont {E.}~\bibnamefont {Lucero}}, \bibinfo {author} {\bibfnamefont
  {R.}~\bibnamefont {McDermott}}, \bibinfo {author} {\bibfnamefont
  {M.}~\bibnamefont {Neeley}}, \bibinfo {author} {\bibfnamefont {E.~M.}\
  \bibnamefont {Weig}}, \bibinfo {author} {\bibfnamefont {A.~N.}\ \bibnamefont
  {Cleland}}, \ and\ \bibinfo {author} {\bibfnamefont {J.~M.}\ \bibnamefont
  {Martinis}},\ }\href {\doibase 10.1126/science.1130886} {\bibfield  {journal}
  {\bibinfo  {journal} {Science (New York, N.Y.)}\ }\textbf {\bibinfo {volume}
  {313}},\ \bibinfo {pages} {1423} (\bibinfo {year} {2006})}\BibitemShut
  {NoStop}%
\bibitem [{\citenamefont {Reed}\ \emph {et~al.}(2010)\citenamefont {Reed},
  \citenamefont {DiCarlo}, \citenamefont {Johnson}, \citenamefont {Sun},
  \citenamefont {Schuster}, \citenamefont {Frunzio},\ and\ \citenamefont
  {Schoelkopf}}]{Reed2010}%
  \BibitemOpen
  \bibfield  {author} {\bibinfo {author} {\bibfnamefont {M.~D.}\ \bibnamefont
  {Reed}}, \bibinfo {author} {\bibfnamefont {L.}~\bibnamefont {DiCarlo}},
  \bibinfo {author} {\bibfnamefont {B.~R.}\ \bibnamefont {Johnson}}, \bibinfo
  {author} {\bibfnamefont {L.}~\bibnamefont {Sun}}, \bibinfo {author}
  {\bibfnamefont {D.~I.}\ \bibnamefont {Schuster}}, \bibinfo {author}
  {\bibfnamefont {L.}~\bibnamefont {Frunzio}}, \ and\ \bibinfo {author}
  {\bibfnamefont {R.~J.}\ \bibnamefont {Schoelkopf}},\ }\href {\doibase
  10.1103/PhysRevLett.105.173601} {\bibfield  {journal} {\bibinfo  {journal}
  {Physical Review Letters}\ }\textbf {\bibinfo {volume} {105}},\ \bibinfo
  {pages} {173601} (\bibinfo {year} {2010})}\BibitemShut {NoStop}%
\bibitem [{\citenamefont {Aron}\ \emph
  {et~al.}(2014{\natexlab{b}})\citenamefont {Aron}, \citenamefont {Kulkarni},\
  and\ \citenamefont {T\"{u}reci}}]{Aron2014b}%
  \BibitemOpen
  \bibfield  {author} {\bibinfo {author} {\bibfnamefont {C.}~\bibnamefont
  {Aron}}, \bibinfo {author} {\bibfnamefont {M.}~\bibnamefont {Kulkarni}}, \
  and\ \bibinfo {author} {\bibfnamefont {H.~E.}\ \bibnamefont {T\"{u}reci}},\
  }\href {http://arxiv.org/abs/1412.8477} {\  (\bibinfo {year}
  {2014}{\natexlab{b}})},\ \Eprint {http://arxiv.org/abs/1412.8477}
  {arXiv:1412.8477} \BibitemShut {NoStop}%
\end{thebibliography}%


%

\end{document}